\documentclass[conference]{IEEEtran}
\IEEEoverridecommandlockouts

\usepackage{amsmath,amssymb,amsfonts}

\usepackage{braket}

\usepackage{algorithmic}
\usepackage{graphicx}
\usepackage{tikz}
\usetikzlibrary{shadows.blur}
\usetikzlibrary{shapes.symbols}

\tikzset{treenode/.style={black!70!blue!90, draw=gray, shade, top color=gray!25!blue!25,
      bottom color=gray!10!blue!10,fill=gray!10!blue!10, thick, minimum size=0.5cm, rounded corners, drop shadow,  shadow xshift = -5cm
}}

\usepackage{textcomp}
\usepackage{xcolor}

\usepackage{placeins}
\usepackage{adjustbox}

\usepackage{enumerate}

\def\BibTeX{{\rm B\kern-.05em{\sc i\kern-.025em b}\kern-.08em
    T\kern-.1667em\lower.7ex\hbox{E}\kern-.125emX}}
\usepackage{cite}
\usepackage{url}

\begin{document}

\title{Bayesian Optimization for Repeater Protocols}

\author{
    \IEEEauthorblockN{
        Lorenzo La Corte\IEEEauthorrefmark{1}\IEEEauthorrefmark{2}, 
        Kenneth Goodenough\IEEEauthorrefmark{3}, 
        Ananda G. Maity\IEEEauthorrefmark{1}, 
        Siddhartha Santra\IEEEauthorrefmark{4} and 
        David Elkouss\IEEEauthorrefmark{1}
    }
    
    \IEEEauthorblockA{
        \IEEEauthorrefmark{1}
        \textit{Networked Quantum Devices Unit},
        \textit{Okinawa Institute of Science and Technology Graduate University},
        \textit{Okinawa, Japan}
    }
    \IEEEauthorblockA{
        \IEEEauthorrefmark{2}
        \textit{DIBRIS},
        \textit{University of Genoa},
        \textit{Genoa, Italy}
    }
    
    \IEEEauthorblockA{
        \IEEEauthorrefmark{3}
        \textit{Manning College of Information and Computer Sciences},
        \textit{University of Massachusetts Amherst},
        \textit{Amherst, USA}
    }
    
    \IEEEauthorblockA{
        \IEEEauthorrefmark{4}
        \textit{Department of Physics and CoE-QUICST},
        \textit{Indian Institute of Technology, Bombay},
        \textit{Powai, Maharashtra, India}
    }
}
\maketitle
\begin{abstract}
Efficiently distributing secret keys over long distances remains a critical challenge in the development of quantum networks. 
``First-generation" quantum repeater chains distribute entanglement 
by executing protocols composed of probabilistic entanglement generation, swapping and distillation operations. 
However, finding the protocol that maximizes the secret-key rate is difficult for two reasons. First, calculating the secret-key rate for a given protocol is non-trivial due to experimental imperfections and the probabilistic nature of the operations. Second, the protocol space rapidly grows with the number of nodes, and lacks any clear structure for efficient exploration. To address the first challenge, we build upon the efficient machinery developed by Li et al.~\cite{Li_2021} and we extend it, enabling numerical calculation of the secret-key rate for heterogeneous repeater chains with an arbitrary number of nodes.
For navigating the large, unstructured space of repeater protocols, we implement a Bayesian optimization algorithm, which we find consistently returns the optimal result. Whenever comparisons are feasible, we validate its accuracy against results obtained through brute-force methods.
Further, we use our framework to extract insight on how to maximize the efficiency of repeater protocols across varying node configurations and hardware conditions. 
Our results highlight the effectiveness of Bayesian optimization in exploring the potential of near-term quantum repeater chains.
\end{abstract}

\begin{IEEEkeywords}
Quantum network, Bayesian optimization, entanglement distillation.
\end{IEEEkeywords}

\section{Introduction}\label{sec:introduction}
High-quality entanglement among distant parties is a crucial resource for distributed information processing and secure communication, enabling quantum key distribution~\cite{Bennett_1984,Ekert_1991}, improved interferometry for telescopes~\cite{Gottesman_2012}, enhanced clock synchronization~\cite{Kmr_2014}, distributed quantum computing~\cite{Buhrman_2003}, and ultimately the development of quantum networks~\cite{Wehner_2018,Azuma_2023}.
However, establishing entanglement between distant nodes is a challenging task, as the direct transmission of photons is hindered by an exponential loss in physical channels.

Quantum repeater stations, placed between the end nodes, have been proposed as one possible solution to overcome these challenges~\cite{Azuma_2023}. Instead of direct transmission, entanglement is created over segments of shorter lengths and then transformed into a single long-distance one using \emph{entanglement swapping}.
Protocols for entanglement distribution leverage entanglement swapping and distillation to generate long-distance entanglement while maintaining its quality high enough for practical use. 
However, these processes are prone to failure and can, in general, be affected by noise and imperfections of the system, making a real implementation of quantum repeater protocols challenging in practice.

While entanglement swapping extends entanglement over longer distances, it degrades the quality of the output states if even slight noise affects the states or measurements. In contrast, entanglement distillation boosts state quality (fidelity to a Bell state) but adds waiting time, as it requires multiple copies (treating entanglement generation as a sequential process). 
Furthermore, the probability of success of a distillation attempt depends on the quality of the input links, which makes previous stages of the protocol crucial for the amount of time needed for the distribution of entanglement. Schemes based on heralded entanglement generation and (probabilistic) entanglement distillation (but no quantum error correction) are known as ``first-generation" quantum repeaters~\cite{Munro_2015}.
For explicit classification of quantum repeater protocols into distinct generations, see Ref.~\cite{Muralidharan_2016}.

Focusing on ``first-generation" quantum repeater schemes, here we aim to efficiently distribute entanglement over long distances by designing a protocol that, under given hardware conditions, optimally sequences swapping and distillation.

The structure of the article is as follows. In Section~\ref{sec:preliminaries}, we discuss the model introduced in~\cite{Li_2021} to simulate quantum repeater chains, and we extend it by relaxing its assumptions on the number of nodes and the homogeneity of the hardware in the chain.
In Section~\ref{sec:protocols}, we introduce the notation for a single protocol and define the notion of protocol space.
We explore large spaces of repeater protocols for chains where assumptions on the number and hardware quality of nodes have been relaxed.
In Section~\ref{sec:optimization}, we introduce a method for efficiently investigating the protocol spaces, followed by a discussion of the results in Section~\ref{sec:results}.

\subsection{Our contribution}
We consider a repeater chain distributing entanglement at a target distance, and a set of values determining the quality of its underlying hardware (e.g. the probability of generating elementary entanglement between two consecutive nodes). We consider the set (space) of all protocols achieving an end-to-end link (connecting the two endpoints in the chain), and we search for the most efficient one.
In particular, we efficiently simulate a single quantum repeater protocol by extending and refining the model from Ref.~\cite{Li_2021}, and develop a method to analyze the space of protocols distributing entanglement in the given chain.
We discuss optimal entanglement swapping strategies and the trade-off between improved end-to-end fidelities through distillation and the additional waiting times.

Specifically, we address the following questions. Given the distance at which we desire to distribute entanglement, we search for the most efficient number of intermediate stations (repeaters) between the end parties. Also, having set the total number $N$ of nodes in the chain (considering both end parties and repeaters), we seek the most effective method for swapping the entangled links.

We then aim to identify hardware conditions where distillation is beneficial or detrimental. In regimes where distillation is advantageous, we also determine the optimal number of distillations and the most efficient way of arranging them with swapping throughout the protocol. We study whether it is better to perform entanglement distillation before swapping or to swap immediately after generating the entangled link.

While addressing these questions, we give the following contributions. (i) we build on the algorithm from~\cite{Li_2021}, which evaluates repeater chains with $N = 2^n + 1$ nodes (with $n$ being a positive integer), extending it to simulate chains with any number of nodes. We then introduce another extension to simulate heterogeneous chains, where each node or link can have different qualities. (ii) Introducing the notation of \emph{protocol space}, defined as the set of all the possible protocols for distributing entanglement across a chain of $N$ nodes, we implement a Bayesian optimization algorithm to investigate a vast space of repeater protocols in a reasonable time.

\subsection{Related work}
While extensive research exists on quantum repeater chains (see~\cite{Azuma_2023} for a detailed review), finding the optimal strategies for quantum repeater protocols involving distillation remains largely unexplored.

Simple entanglement swapping (ES) strategies perform ES in a \emph{nested} fashion~\cite{Briegel_1998}, or as soon as two adjacent entangled links are available (\emph{SWAP-ASAP})~\cite{goodenough2024noise}.
In Ref.~\cite{Inesta_2023}, the authors consider homogeneous chains and conclude that a nested strategy generally reduces the waiting time needed to distribute entanglement.
Studies on optimal policies for ES have been proposed in Ref.~\cite{Inesta_2023, Haldar_2024_RL}, both based on a Markov Decision Process (MDP) for repeater chains, and a learning algorithm solving it. The optimal policies found lead to an improvement in the efficiency of the repeater protocols.
The algorithm aims to choose the most efficient sequence of actions maximizing the quality of the end-to-end entangled link and minimizing the time needed to achieve it. 
However, they use a simplified model to describe the state of the chain. 
They model decoherence using the \emph{age} of an entangled link, i.e. the count of time steps from when it is generated. Fidelity is derived from the age, and a link is generated again if its age exceeds a certain threshold. They assume Pauli noise, using the "addition rule" to compute the age for a link in output of a swap operation as the sum of the ages of the input links.  
We employ a more sophisticated model by building upon the framework developed in Ref.~\cite{Li_2021}, discussed in more detail in Section \ref{sec:model}, which, for instance, includes time-dependent exponential decay decoherence. It also treats both the waiting time and the end-to-end state's entanglement quality as random variables. 
For each instant of time, distributions of these variables are computed.
Thus, in contrast to~\cite{Haldar_2024_RL}, we get the complete view of the efficiency of all the possible sequences of actions performed to distribute entanglement.

In addition, Ref.~\cite{Li_2021} also examines the impact of cut-off conditions on the links of the chain. An example of such a condition is a limit on the maximum storage time of a link in an imperfect memory, after which it is discarded and regenerated.
Unlike~\cite{Inesta_2023, Haldar_2024_RL, Li_2021} we do not take cut-off conditions into account, thus considering a smaller set of possible actions to be performed on the repeater chain.

Finally, we refer to studies focusing on optimal strategies for entanglement distillation.
Ref.~\cite{VanMeter_2008, Nagayama_2021} consistently find that performing entanglement distillation at link-level (immediately after entanglement generation) is essential for maintaining high-fidelity quantum states over long distances. In particular, Ref.~\cite{VanMeter_2008} finds that link-level distillation is particularly effective when the initial state fidelities are low, improving the success probability of the following swapping or distillation operations. Similarly, Ref.~\cite{Nagayama_2021} shows that, by addressing errors locally, link-level distillation prevents error accumulation, improving the overall efficiency of the repeater protocol.
Ref.~\cite{Haldar_2024_Quasi} examines the efficiency of distillation-based policies and shows that performances depend heavily on hardware parameters, such as the quality of elementary links.
They find that distillation is often detrimental when elementary links have high fidelity, as it increases waiting times, but becomes beneficial for elementary links of low fidelity.

\section{Preliminaries}\label{sec:preliminaries}

\subsection{Class of Repeater Protocols Considered}\label{sec:class}

We consider protocols that distribute entanglement through three quantum network primitives: entanglement generation, swapping and distillation.
These primitives operate as stochastic heralded processes i.e., they succeed probabilistically and the outcomes are communicated to the parties via classical messages. We leave out the study of cut-offs. See~\cite{Li_2021,Inesta_2023, Haldar_2024_RL} for recent progress in the optimization of cut-offs. 

All three processes have (sets of) entangled links as in- and output. 
Entanglement generation acts upon a pair of adjacent nodes and aims to produce an entangled link between them. It takes no input and outputs a single link in case of success.
These freshly generated entangled pairs are termed elementary links or link-level entanglement. 
Entanglement swapping aims to extend entanglement to larger distances by taking two adjacent links as inputs and combining them into a single longer link of reduced quality.
Entanglement distillation improves the quality of the entangled state, taking two links between the same pair of nodes as input and probabilistically producing a higher-quality link.
In case of failure of any of these processes, all the links involved are lost and have to be generated again. Since all these probabilistic processes are stacked on top of each other, they all have to succeed for the protocol to produce a final end-to-end entangled link. 

\begin{figure}[tb]
  \centering
  \includegraphics[width=\linewidth, trim={0 0.05cm -1cm 0.05cm},clip]{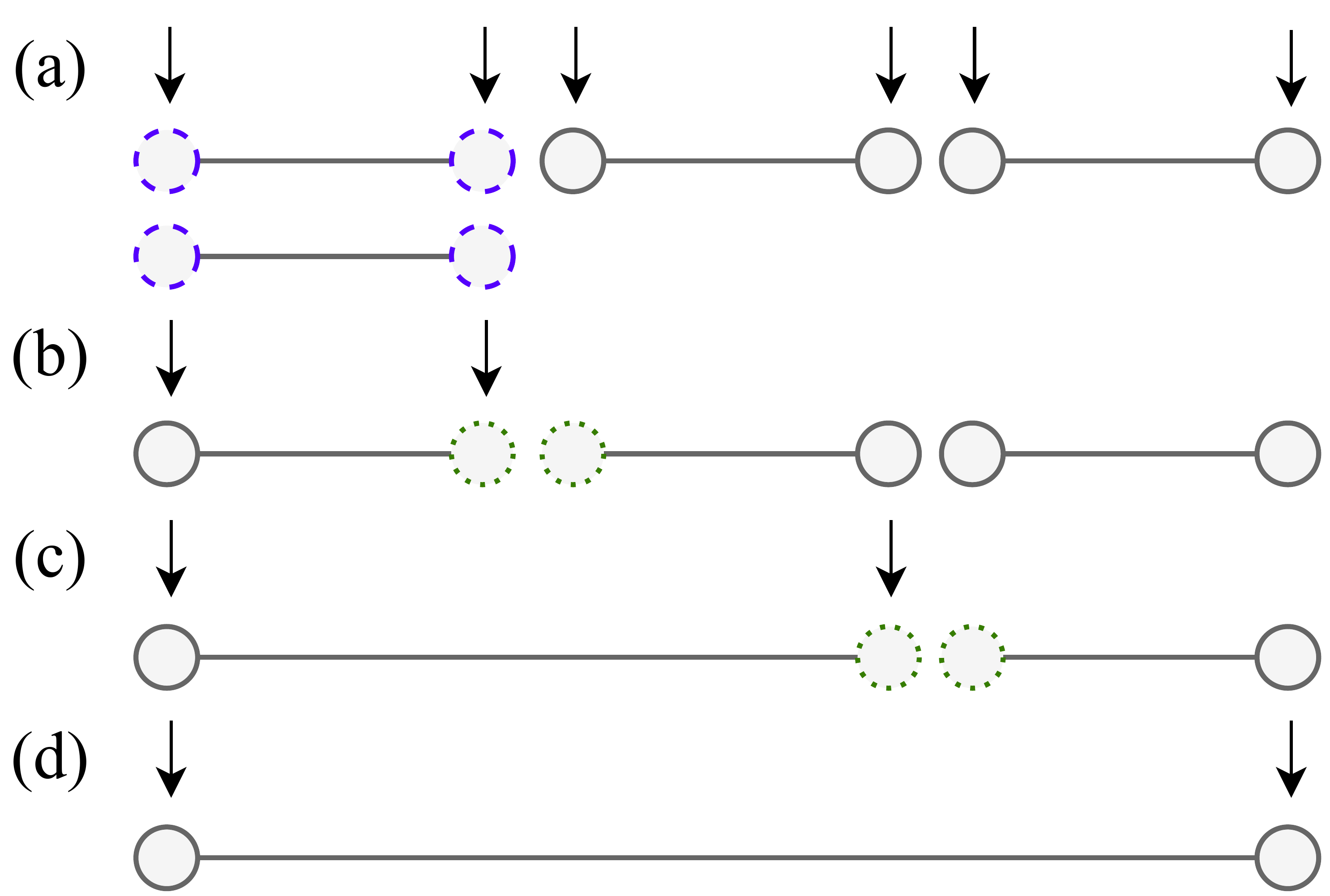}
  \caption{Visual representation of a repeater protocol involving entanglement distillation and swapping. (a) All links are probabilistically generated. (b) Entanglement distillation is performed on the leftmost couple of nodes: in case of success, the two entangled links are distilled into a single one of higher quality. (c) Adjacent entangled links are probabilistically swapped, and then (d) the output link is swapped again to have end-to-end entanglement.}
\label{fig:chain_example}
\end{figure}

Fig.~\ref{fig:chain_example} shows a visual representation of an example of a protocol. 
All processes (arrows) are probabilistic. In case of failure, the input links are lost and need to be regenerated.
The protocol is performed as follows. Two entangled links are generated between the leftmost pair of nodes, while only one link is created for the remaining segments. 
The two left links are distilled into a higher-quality link, which is then swapped with the adjacent link to form a longer connection. Finally, a last swapping operation establishes entanglement between the chain's endpoints, completing the protocol.

\subsection{Model}\label{sec:model}

We simulate repeater chains using the model and algorithm developed in~\cite{Li_2021}.
We consider a chain having $N$ nodes that are connected by $N-1$ entangled links. The goal is to establish end-to-end entanglement between its two extreme nodes using a sequence of distillation and swapping. Unlike the work from~\cite{Li_2021}, we do not assume that the chains are \emph{symmetric} i.e., the number of links need not be a power of two. Furthermore, we do not assume that the chains are \emph{homogeneous}. We allow each link $i$ and each node $j$ to have different values for their entanglement generation probability $p_\text{gen}^{(i)}$, initial link quality $w_0^{(i)}$ and memory coherence time $t_\text{coh}^{(j)}$.
For simplicity, we assume that the probability of success for entanglement swapping $p_\text{swap}$ is always the same. 

We note that the distance between two nodes in the chain impacts the probability of generation $p_\text{gen}$ and the initial quality $w_0$ of the entangled link.
Thus, considering different starting probabilities of entanglement generation and initial qualities is a proxy to simulate chains of unevenly spaced nodes. However, this is an approximation, as we do not consider delays in classical communication.

The model aims to incorporate two figures of merit: (i) the time required for distributing entanglement and (ii) the quality of the final entangled link. Since entanglement generation, swapping and distillation are all stochastic processes, they are treated as random variables from which probability distributions are derived.

We assume that the entangled links produced at each step of the protocol are Werner states: 
\begin{equation}\label{eq:WernerStateLi}
    \rho(w) = w \ket{\Phi^+}\bra{\Phi^+} + (1 - w) \frac{\mathbb{I}}{4} .
\end{equation}
Here $w \in [0,1]$ is the Werner parameter which is also related to the fidelity $F$ (with $\ket{\Phi^+}\bra{\Phi^+}$) as $F = (1 + 3w)/4$. We consider Werner states because they naturally arise in experimental setups, and any two-qubit state can be transformed into it using a standard twirling argument \cite{bennett1996mixed}. 

We treat each heralded attempt of entanglement generation as independent, with a duration given by:
\begin{equation}
    t_\text{unit} = L_0 / c
\end{equation}
where $L_0$ is the distance between the nodes in the chain, and $c$ is the speed of light. This is also the discrete time unit employed by the model.
When generated, an entangled pair has an initial quality $w_0$ that depends on the underlying hardware, and it is thus an input parameter for the model and the algorithm.

After generation, pairs are stored in imperfect memories for a time $t$ and decohere as:
\begin{equation}\label{eq:decoherence_general}
    w^{\prime} = w ~ e^{-t/t_\text{coh}^A} e^{-t/t_\text{coh}^B}
\end{equation}
where $t_\text{coh}^A$ and $t_\text{coh}^B$ are the coherence times of the involved nodes' memories. Unlike~\cite{Li_2021}, we do not assume $t_\text{coh}^A=t_\text{coh}^B$.

We investigate the effect of decoherence in probabilistic chains. We assume that entanglement generation and swapping succeed with probabilities $p_\text{gen}$ and $p_\text{swap}$, respectively. For entanglement distillation, we use the BBPSW protocol from Bennett \textit{et al.} \cite{Bennett_1996} where the probability of success for one attempt of a single round of distillation is given by:
\begin{equation}
    p_\text{dist} = \frac{1+w_A^{\prime} w_B^{\prime}}{2},
\end{equation}
where $w_A$ and $w_B$ are the Werner parameters of the links, and the primed notation denotes the Werner parameter with decay in~\eqref{eq:decoherence_general} applied to the link that waits until the other is ready.

Since a successful attempt at swapping or distillation always outputs Werner states, the Werner parameters of the output links generally depend on the input links. For swapping and distillation, these are respectively given by:
\begin{equation}
    w_\text{swap} = w_A^{\prime} w_B^{\prime},\quad
    w_\text{dist} = \frac{w_A^{\prime}+w_B^{\prime}+ 4 w_A^{\prime} w_B^{\prime}}{6 p_\text{dist}}
\end{equation}
where decoherence is taken into account (prime notation) for one of the two input links, as described above. 

Using these primitives and the four hardware parameters ($p_\text{gen}$, $p_\text{swap}$, $w_0$, $t_\text{coh}$), expressions for the waiting time needed and the Werner parameter of the end-to-end link are derived and then converted into an efficient algorithm~\cite{Li_2021}.
This algorithm implements these expressions by truncating the computation of the distribution at a time $t_\text{trunc}$, which can be set sufficiently high to cover the $1-\epsilon$ region of the cumulative distribution.

\section{Repeater Protocols}\label{sec:protocols}

\begin{figure}[tb]
  \centering
  \begin{tikzpicture}[
    level distance=0.8cm,
    sibling distance=4cm,
    level 2/.style={sibling distance=2cm},
    every node/.style={inner sep=2mm}
  ]
    \node[treenode] {2}
        child {node[treenode] {0}
            child {node[treenode] {1}}
            child {node[treenode] {1}}
        }
        child {node[treenode] {1}
        };
  \end{tikzpicture}
  \caption{Visual representation of a single repeater protocol, for a chain of $N=4$ nodes. Each vertex of the tree represents an entangled link in the chain. Leaves of the tree represent freshly generated links. In this case, we have $3$ leaves, which are all distilled once, as shown on their label. When two entangled links (vertices) are swapped, the output link is represented in the tree as their parent. The root represents the end-to-end link, on which two rounds of entanglement distillation are performed.}
  \label{fig:binary_tree_notation}
\end{figure}
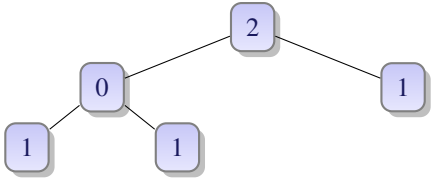

We represent a protocol as a binary tree, where each vertex represents a link, and their label indicates the number of distillation rounds performed on that link, as shown in Fig.~\ref{fig:binary_tree_notation}. 
The leaves of the binary tree represent the links joining nodes after entanglement generation. An instance of entanglement swapping produces a new entangled link, represented as the parent of the two input links.

We define a \emph{protocol space} as the set of all possible protocols for a chain of $N$ nodes with up to a maximum of $\beta$ rounds of distillation performed on each entangled link at any level.
The size of the protocol space is determined by the number of all possible sequences of operations for distributing entanglement, considering the restrictions imposed.

The structure of a binary tree represents the order of the operations performed.
Thus, the number of valid possible sequences of entanglement swapping operations is given by the number of all the possible binary trees with $v = 2N - 3$ vertices. Equivalently, these are all possible binary trees with $N-1$ leaves, which is given by the $(N-2)$'th Catalan number $C(N-2)$, where
\begin{equation}
  C(n) = \frac{1}{n+1} \binom{2n}{n} .
\end{equation}

We account for all possible numbers of distillation rounds for any link at any level in the chain (i.e., for all values of $k \in [0, \beta]$ to label each tree vertex).
Thus, for $N$ nodes, the size of the protocol space is given by:
\begin{equation}
    |P| = C(N-2)(\beta+1)^v\label{eq:size_of_protocolspace} .
\end{equation}

\section{Efficient Investigation of Protocol Spaces}\label{sec:optimization}

%

\subsection{Figure of Merit}\label{sec:figure_of_merit}

In our numerical study, we employ the secret-key rate (SKR) of the BB84 protocol~\cite{Bennett_1984} as the figure of merit to assess the performance of a repeater protocol. It quantifies the rate at which secret bits can be generated between the two end parties, and it is computed by dividing the secret-key fraction---a function of the quality of the entanglement---by the average waiting time.
The secret-key rate is a practical measure of entanglement usefulness~\cite{Vardoyan_2022} as it assesses both the quality of the end-to-end link and the efficiency of the repeater protocol in producing entangled pairs.

\subsection{Bayesian Optimization}\label{sec:BayesianAlgorithm}
We aim to find the repeater protocol that maximizes the secret-key rate. To efficiently explore the space we leverage the \emph{Bayesian optimization} algorithm from the scikit-optimize library~\cite{skopt}. Bayesian optimization is a strategy to globally optimize black-box functions $f: A\rightarrow \mathbb{R}$, where $A \subseteq \mathbb{R}^d$ for some $d$ (see~\cite{Frazier_2018} for more details). 

Let us first consider a naive application of Bayesian optimization to our setting. The black-box function would be the secret-key rate, computed as a function of the used protocol. Since the function $f$ takes as input elements of $\mathbb{R}^d$, each protocol would be represented by a tuple. 
Since the protocol space is discrete, it could be mapped bijectively to the set $A=\lbrace{1, 2, \ldots, C(N-2) \left(\beta+1\right)^v\rbrace}$ (which is a subset of $\mathbb{R}^1$), derived from a chosen ordering of the protocols.

We have found that in practice the above approach performs poorly, and thus use a different set-up. Instead, a tuple $\left(\gamma, \mathcal{K}, \eta, \tau\right)\in \mathbb{R}^4$ encodes now a \emph{probability distribution} over protocols. The four parameters describe qualitative features of the sampled protocols. For example, if $\mathcal{K}=0$, the strategy is to avoid distillation (see more details below). Now, to compute our `function' $f$ with input $\left(\gamma, \mathcal{K}, \eta, \tau\right)$, we sample a protocol from the associated probability distribution.
We found that this approach significantly outperforms the above naive approach.

We now make explicit the probability distribution associated with the tuple $\left(\gamma, \mathcal{K}, \eta, \tau\right)$:
\begin{itemize}
    \item $\gamma \in [0,1]$ represents the degree of symmetry, i.e. how balanced the binary tree is, defined here as \emph{symmetricity},
    \item $\mathcal{K} \in \lbrace{0, 1, \ldots, v \beta\rbrace}$ is the total rounds of distillations performed on the entangled links at any level of the chain,
    \item $\eta \in [-1, 1]$ and $\tau \in [0,1]$ are used to respectively derive the mean and variance of the distribution, from which we sample how the $\mathcal{K}$ rounds of distillations are distributed among the $v$ vertices of the tree.  
\end{itemize}

Starting from these input parameters, the encoding algorithm works as follows.
First, considering all the possible binary trees with $v$ vertices, it generates the set of all possible sequences of swaps $P_\text{swap}$. It orders the trees from the least symmetric to the most symmetric by their \emph{symmetricity score} $\mathcal{S}$. 
The symmetricity score $\mathcal{S}$ of a binary tree shape is given by:
\begin{equation}
    \mathcal{S} = 1 - \frac{\mathrm{Var}(\mathcal{L)}}{\mathcal{L}_{\max}}
\end{equation}
where $\mathrm{Var}(\mathcal{L})$ is the variance of the depths of the leaves, defined as:
\begin{equation}
    \mathrm{Var}(\mathcal{L}) = \frac{1}{n} \sum_{i=1}^{n} (l_i - \bar{l})^2
\end{equation}
with $l_i$ representing the depth of the $i$-th leaf, and $\bar{l}$ and $\mathcal{L}_{\max}$ being, respectively, the mean and the maximum depth of the leaves in the tree.

The algorithm selects the tree in the ordering closest to the value $\gamma \times |P_\text{swap}|$. We note that $\gamma = 0$ corresponds to the least symmetric tree and $\gamma = 1$ with the maximally symmetric one. %

For the selected sequence, we get a shape of a binary tree with $v$ vertices, in which we aim to distribute the $\mathcal{K}$ total rounds of distillation.
We can visualize the rounds of distillation for the $v$ vertices as a tuple:
\begin{equation}
    (\kappa_1, \kappa_2, \ldots, \kappa_v) \quad \text{with} \quad \sum_{i=1}^{v} \kappa_i = \mathcal{K} ,
\end{equation}
where the leftmost values correspond to the rounds of distillation assigned to the leaves of the binary tree, i.e., the elementary links, whereas the rightmost value corresponds to the rounds assigned to the tree root, i.e., the end-to-end link.
We can sample the values for this tuple from a normal distribution $\mathcal{N}(\mu, \sigma)$, where
\begin{align}
  \mu = \frac{(\eta + 1) v}{2},\quad\sigma = \tau v
\end{align}
and restrict the maximum value of $\kappa_i$ to $\beta$.
If $\eta = -1$, the encoding algorithm selects sequences that distill at the link-level. On the other hand, if $\eta = 1$, the algorithm selects sequences that perform distillation as the final step of the protocol (end-to-end level).
The parameter $\tau$ determines how strong the strategy given by the parameter $\eta$ is. The higher the value of $\tau$, the more random the choice of protocol becomes.

\subsection{Complexity}

A naive brute-force approach would require us to evaluate the secret-key rate for each protocol in the protocol space (see Eq.~\eqref{eq:size_of_protocolspace}). The number of \emph{shots}, i.e.~the number of protocols evaluated for the Bayesian optimization, is given as input to the algorithm and should be smaller than $\left|P\right|$. This parameter can be set to a much lower value with respect to the cardinality of the space of the protocols, while still converging to the optimal solution in practice.

The cost of one shot is given by the complexity of the algorithm developed in~\cite{Li_2021}, which is quadratic in the truncation time for the simulation.
A more efficient version of the algorithm in~\cite{Li_2021} with quasilinear complexity $\mathcal{O}(t_\text{trunc} \log t_\text{trunc})$ is possible if the success probability and Werner parameter of swapping and distillation admit a factorization in terms of two sets of real functions $\{f^{(i)}\},\{g^{(i)}\}$ where $f^{(i)}$ functions and $g^{(i)}$ functions depend respectively only on the times $t_1$ and $t_2$ at which each link is ready:
\begin{equation}\label{eq:factorization}
    \sum_i f^{(i)} (t_1)  g^{(i)} (t_2).
\end{equation}
In the following, we show that such a factorization is also possible for the extension to heterogeneous chains.

We follow alongside the steps described in~\cite{Li_2021}, where the complexity reduction is achieved for homogeneous chains, with fixed coherence times for all the nodes. 
In heterogeneous chains, each node memory has a different coherence time. 
A link thus has the two memory qubits waiting in two different nodes with memory decay factors:
$\exp{(-|t_1 - t_2| / t_\text{coh}^A)}$ and $\exp{(-|t_1 - t_2|/t_\text{coh}^B)}$. The overall decay factor is given by the multiplication of the two factors: 
\begin{equation}
    \exp{\left(- \frac{|t_1 - t_2|}{t_\text{coh}^A}-\frac{|t_1 - t_2|}{t_\text{coh}^B}\right)}.
\end{equation}

Now, suppose $t_1 \geq t_2$. The overall decay factor can also be written in the following way: %
\begin{equation}
    \exp{\left(- \frac{t_1}{t_\text{coh}^A} - \frac{t_1}{t_\text{coh}^B}\right)} 
    \exp{\left(+ \frac{t_2}{t_\text{coh}^A} + \frac{t_2}{t_\text{coh}^B}\right)}.
\end{equation}
By noting that the first exponential only depends on $t_1$ and the second one on $t_2$, this factorization directly allows constructing $f$ depending only on $t_1$ and $g$ depending only on $t_2$ as described in Eq.~\eqref{eq:factorization}. A similar argument holds if $t_1\leq t_2$.

%

\section{Numerical Results}\label{sec:results}
In this section, we optimize the search for the repeater protocol in the space achieving the maximum secret-key rate.

We begin by focusing on homogeneous chains, using the Bayesian optimization algorithm presented in Section~\ref{sec:BayesianAlgorithm} to study the impact of entanglement distillation and of the symmetricity of entanglement swapping sequences (binary tree shapes).
Table~\ref{tab:hardware_regimes} outlines the hardware regimes considered. Here $L_0$ is the distance between the nodes of the chain, $p_\text{gen}$ is the probability of success for the generation of an elementary link, and $w_0$ is its initial Werner parameter. Realistic values for these hardware parameters are chosen from Da Silva \textit{et al.}~\cite{DaSilva_2021}, which evaluates the performance of a quantum network using data from SURF, a cooperative of Dutch education and research institutions~\cite{surf}. Since we first simulate homogeneous chains, $t_\text{coh}$ represents the joint coherence time of the two node memories involved in a link, given in units of time of the simulation. The entanglement swapping success probability is fixed at $p_\text{swap} = 0.85$ for all scenarios.
\begin{table}[b]
    \caption{Hardware regimes considered for the simulations.}
    \begin{center}
    \begin{tabular}{|c|c|c|c|c|}
        \hline
        \textbf{Scenario} & \multicolumn{4}{|c|}{\textbf{Hardware Parameters}} \\
        \cline{2-5}
        & $L_0$ (km) & $t_\text{\text{coh}}$ ($L_0/c$) & $w_0$ & $p_\text{gen}$ \\
        \hline
        A & 200  & $3.6 \times 10^5$ & 0.360 & $9.6 \times 10^{-8}$ \\
        \hline
        B & 100  & $7.2 \times 10^5$ & 0.867 & $1.5 \times 10^{-5}$ \\
        \hline
        C & 50   & $1.4 \times 10^6$ & 0.952 & $9.2 \times 10^{-4}$ \\
        \hline
        D & 20   & $3.6 \times 10^6$ & 0.958 & $2.6 \times 10^{-3}$ \\
        \hline
    \end{tabular}
    \end{center}
    \label{tab:hardware_regimes}
\end{table}
For repeater protocols in homogeneous hardware settings, we show that the optimal protocol is the most symmetric one, represented by a maximally balanced binary tree.
We also show that distillation is either detrimental if the quality of generated links is high, or optimal if performed right after entanglement generation.

We then extend our study to heterogeneous protocols, where general insights are less straightforward. We examine hardware regimes where nodes and links of the chain have different qualities. We show that to maximize the efficiency of a protocol, it is essential to balance the waiting times required for both links involved in swap operations to be ready.

In both settings, we investigate spaces of repeater protocols described in Section~\ref{sec:protocols}. Using our open-source implementation~\cite{repo} of the optimization algorithm introduced in Section~\ref{sec:BayesianAlgorithm} we conduct the study on the university computational cluster~\cite{deigo}. When possible, we also validate the results by comparing them with a brute-force algorithm that evaluates all possible protocols in the space.

All the simulations shown have sufficient truncation time to cover the $99\%$ of the distributions for the waiting time and the quality of the end-to-end entangled link, from which we compute the secret-key rate (see Section~\ref{sec:figure_of_merit}).

\subsection{Entanglement Distribution at a Given Distance}\label{sec:results_distribution}

We present the optimization results for spaces of protocols that distribute entanglement at a distance of $200$km. We examine four settings (Table~\ref{tab:hardware_regimes}), each with a different internode distance and set of hardware parameters.
The repeater chains considered are homogeneous in their hardware and generally asymmetric (the number of elementary links is not limited to a power of two).
The number of intermediate nodes used to extend entanglement varies for each setting, e.g. for scenario~D we need $10$ elementary links, each $20$km long, to connect the two endpoints.
We aim to find the best-performing protocol for each protocol space considered.
In Table~\ref{tab:simulations} we present the results of the simulations using brute-force (top) and Bayesian optimization (bottom) algorithms. For each combination of scenarios and maximum allowed rounds of distillation per link $(\beta)$, we present the time taken, the maximum secret-key rate achieved, and the size (cardinality) of the space.
Each Bayesian optimization simulation performs $100$ shots (see Section~\ref{sec:BayesianAlgorithm}). We do not perform the optimization for scenarios A and B, for which the space cardinality is less than the number of shots. We also do not report results if brute-forcing the whole space is not feasible in a reasonable time.

\begin{table}[t]
    \caption{Time taken and maximum secret-key rate (SKR) found by brute-force (top) and Bayesian optimization (bottom) simulations of the scenarios of Table~\ref{tab:hardware_regimes}, considering a maximum of $\beta$ distillation rounds allowed for each link.}
    \begin{center}
    \begin{tabular}{|c|c|c|c|}
        \hline
        \multicolumn{4}{|c|}{\textbf{Brute-force Simulations}} \\
        \hline
        \textbf{\textit{Scenario}} & \textbf{\textit{Space Cardinality}} & \textbf{\textit{Time Taken}} & \textbf{\textit{Max. SKR}} \\
        \hline
        A, $\beta=2$ & 3 & 1.5 hours & 0.0 \\
        \hline
        B, $\beta=2$ & 27 & 3 days & 0.0 \\
        \hline
        C, $\beta=1$ & $640$ & 3.5 hours & $7.4 \times 10^{-5}$ \\
        \hline
    \end{tabular}
    \end{center}

    \begin{center}
    \begin{tabular}{|c|c|c|c|}
        \hline
        \multicolumn{4}{|c|}{\textbf{Bayesian Optimization Simulations}} \\
        \hline
        \textbf{\textit{Scenario}} & \textbf{\textit{Space Cardinality}} & \textbf{\textit{Time Taken}} & \textbf{\textit{Max. SKR}} \\
        \hline
        C, $\beta=1$ & $640$ & 24 minutes & $7.4 \times 10^{-5}$ \\
        \hline
        C, $\beta=2$ & $1.09 \times 10^{4}$ & 57 minutes & $7.8 \times 10^{-5}$ \\
        \hline
        D, $\beta=2$ & $5.65 \times 10^{12}$ & 7.5 hours & $7.6 \times 10^{-5}$ \\
        \hline
    \end{tabular}
    \end{center}
    \label{tab:simulations}
\end{table}

\begin{figure}[tb]
    \centering
    \begin{tikzpicture}[
        level distance=0.8cm,
        sibling distance=4cm,
        level 2/.style={sibling distance=2cm},
        level 3/.style={sibling distance=1cm},
        every node/.style={inner sep=1mm},
      ]

      \node at (-2.5, 0) {(a)}
      child {node[treenode] {0}
        child {node[treenode] {1}
          child {node[treenode] {1}}
          child {node[treenode] {1}}}
        child {node[treenode] {1}
          child {node[treenode] {1}}
          child {node[treenode] {1}}};
      };

      \node at (2.5, 0) {(b)}
      child {node[treenode] {0}
        child {node[treenode] {0}
          child {node[treenode] {2}}
          child {node[treenode] {2}}}
        child {node[treenode] {0}
          child {node[treenode] {2}}
          child {node[treenode] {2}}};
      };

    \end{tikzpicture}
    \caption{Optimal protocol found by the Bayesian optimization algorithm for scenario C of Table~\ref{tab:hardware_regimes}, with $N=5$ and $\beta=1$ (a) or $\beta=2$ (b).}
    \label{fig:asymmetric_5_nodes_protocol}
\end{figure}
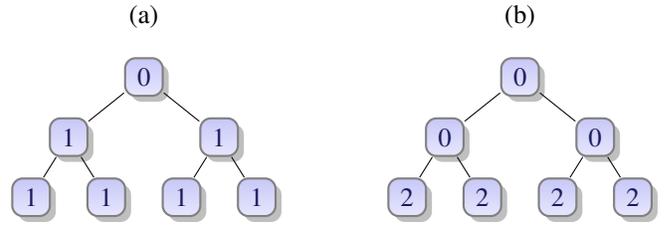

Considering the four scenarios, the optimal protocol is shown in Fig.~\ref{fig:asymmetric_5_nodes_protocol}b.
For the chosen hardware settings, it is the protocol that yields the best secret-key rate extracted from the end-to-end link at $200$ kilometers. Entanglement is generated at $50$ kilometers and extended by parallel instances of entanglement swapping. Distillation is performed two times (the maximum allowed $\beta=2$), only at the link-level.
By allowing one more round of distillation, we improve the result achieved by the protocol presented in Fig.~\ref{fig:asymmetric_5_nodes_protocol}a (with $\beta=1$).

For both the values of $\beta$, the optimal protocol is the maximally symmetric one.
In this hardware-homogeneous chain, it is intuitive that the most symmetric protocol maximizes the secret-key rate, as the waiting time is reduced by performing the entanglement swapping operations in parallel, a result consistent with~\cite{Inesta_2023}.

In both cases, we do not perform distillation on the end-to-end link. This result is also consistent with ~\cite{Haldar_2024_Quasi}, which finds that performing distillation on the end-to-end link introduces waiting times that are detrimental to the secret-key rate.
When we perform entanglement swapping before distillation, the link quality decreases and the distillation process is more likely to fail, as the probability of success of the distillation process depends on the fidelity of the input links.
For this reason, the optimal protocol (Fig.~\ref{fig:asymmetric_5_nodes_protocol}b) performs distillation only after entanglement generation. This is in line with the observations made in~\cite{VanMeter_2008} and~\cite{Nagayama_2021}, which show that performing distillation at the link-level is generally the optimal choice to maximize the efficiency of the protocol.

\begin{figure}[tb]
    \centering
    \begin{tikzpicture}[
        scale=0.5,
        level distance=1.6cm,
        sibling distance=9cm,
        level 2/.style={sibling distance=4.45cm},
        level 3/.style={sibling distance=2.45cm},
        level 4/.style={sibling distance=1.16cm},
        every node/.style={inner sep=1.45mm}
      ]
        \node[treenode] {0}
        child {node[treenode] {0}
            child {node[treenode] {0}
                child {node[treenode] {2}
                    child {node[treenode] {2}}
                    child {node[treenode] {1}}}
                child {node[treenode] {2}
                    child {node[treenode] {2}}
                    child {node[treenode] {2}}}}
            child {node[treenode] {2}
                child {node[treenode] {2}}
                child {node[treenode] {2}}}}
        child {node[treenode] {0}
            child {node[treenode] {2}
                child {node[treenode] {2}}
                child {node[treenode] {2}}}
            child {node[treenode] {2}
                child {node[treenode] {2}}
                child {node[treenode] {2}}}};
    \end{tikzpicture}
    \caption{Optimal protocol found by the Bayesian optimization algorithm for scenario D of Table~\ref{tab:hardware_regimes}, with $N=11$ and $\beta=2$.}
    \label{fig:asymmetric_11_nodes_protocol}
\end{figure}
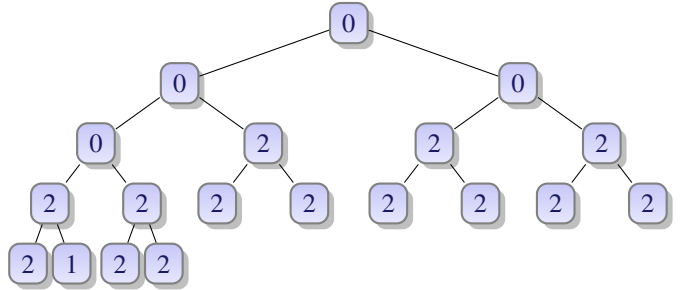

We get similar insights from the results for scenario D, for which the protocol that maximizes the secret-key rate is shown in Fig.~\ref{fig:asymmetric_11_nodes_protocol}.
Although perfect (maximally balanced) trees do not exist in the space for this value of $N$, the optimal shape is one among the most symmetric ones. Again, we see that rounds of distillation are performed at earlier stages of the protocol, and avoided at end-to-end level.

\subsection{Impact of Distillation}\label{sec:results_distillation}

We can investigate further the effect of distillation by considering a repeater chain with fixed hardware parameters, except for the initial Werner parameter $w_0$ of the generated links. Our findings indicate that in low-quality hardware regimes, entanglement distillation enhances the protocol's efficiency.

\begin{figure}[t]
    \centering
    \includegraphics[width=\linewidth, trim={0 0.20cm 0 0},clip]{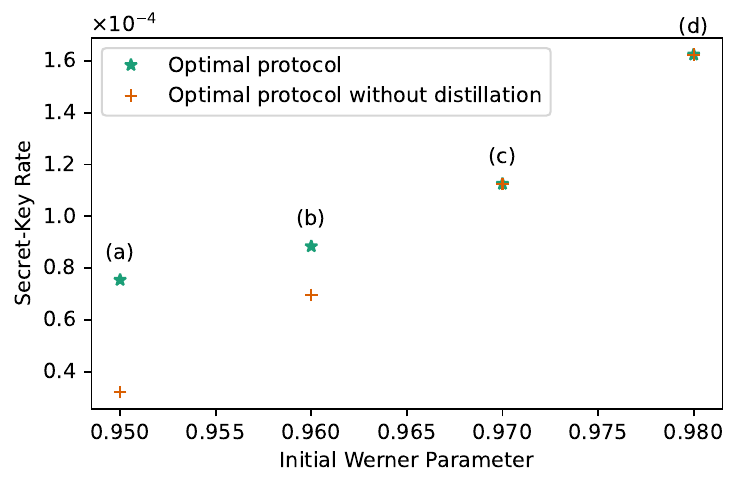}
    \vspace{-5mm}
    \caption{Performance of the optimal protocols found by the Bayesian optimization, in terms of the secret-key rate as a function of the initial Werner parameter $w_0$. For each value of $w_0$ we consider two settings, one allowing a maximum of $\beta=3$ rounds of distillation, and one allowing none. In (a) and (b) the optimal protocols are those in which distillation is performed only at link-level. In (a), entanglement generation of each link is followed by three rounds of distillation, while in (b) one round of distillation is performed. In (c) and (d), the optimal protocols do not perform entanglement distillation.}
    \label{fig:varying_w0}
\end{figure}

We consider a chain of $N=5$ nodes. The initial Werner parameter $w_0$ is set to discrete values ranging from $0.95$ to $0.98$, whereas all the other hardware parameters are set to the values presented for scenario C (Table~\ref{tab:hardware_regimes}).
Fig.~\ref{fig:varying_w0} shows, for each value of $w_0$, the optimal protocols considering only entanglement swapping ($\beta=0$) and also considering a maximum $\beta=3$ of rounds of distillation, which can be applied to each link at any level. All the results are drawn from running the Bayesian optimization algorithm with $500$ shots. The amount of protocols in the space ($|P| = 8.19 \times 10^4$) makes a brute-force simulation unfeasible in a reasonable time. 
For hardware regimes with a high value for $w_0$, distillation is detrimental, as all the protocols involving rounds of distillation yield lower secret-key rates than the one achieved by the optimal protocol involving none.
On the other hand, when the quality of the input links is low ($w_0 \leq 0.96$), distillation enhances the efficiency of the protocol distributing entanglement. In those cases, optimal protocols in terms of the secret-key rate yielded are the ones that perform distillation at early stages.


\subsection{Heterogeneous Chain}\label{sec:results_heterogeneous}

Here, we present the results of the brute-force evaluation of all possible protocols distributing entanglement in a heterogeneous chain consisting of four nodes with different hardware qualities, as detailed in Table~\ref{tab:heterogeneous_hardware_regimes}.
We set different values for the elementary link generation probabilities and their initial Werner parameter. Different times of coherence are set for each node, expressed in units of time of the simulation, as described in Section~\ref{sec:model}.
The probability for success of entanglement swapping $p_\text{swap}$ is constant and set to $0.85$.

\begin{table}[tb]
    \caption{Hardware regimes for nodes and links of a chain based on real-life fiber data~\cite{DaSilva_2021}.}
    \begin{center}
    \begin{tabular}{c c}
        \begin{tabular}{|c|c|}
            \hline
            \textbf{Node} & $t_\text{coh}$ ($L_0/c$) \\
            \hline
            A & $1.08 \times 10^6$ \\
            \hline
            B & $9.50 \times 10^5$ \\
            \hline
            C & $7.20 \times 10^5$ \\
            \hline
            D & $5.60 \times 10^5$ \\
            \hline
        \end{tabular}
        &
        \begin{tabular}{|c|c|c|}
            \hline
            \textbf{Link} & $w_0$ & $p_\text{gen}$ \\
            \hline
            A-B & 0.9577 & 0.25880 \\
            \hline
            B-C & 0.9524 & 0.09187 \\
            \hline
            C-D & 0.9523 & 0.09082 \\
            \hline
        \end{tabular}
    \end{tabular}
    \end{center}
    \label{tab:heterogeneous_hardware_regimes}
\end{table}

\begin{figure}[tb]
    \centering
    \begin{tikzpicture}[
        level distance=0.9cm,
        sibling distance=4cm,
        level 2/.style={sibling distance=2cm},
        every node/.style={inner sep=2mm}
      ]
        \node[treenode] {A-B-C-D, 0}
        child {node[treenode] {A-B-C, 0}
            child {node[treenode] {A-B, 2}}
            child {node[treenode] {B-C, 0}}}
        child {node[treenode] {C-D, 0}};
    \end{tikzpicture}
    \caption{Optimal protocol found by brute-forcing all the protocols for $N=4$, $\beta=2$, and the heterogeneous hardware regime presented in Table~\ref{tab:heterogeneous_hardware_regimes}. For clarity, we add the nodes included in a segment in the vertices labels.}
    \label{fig:heterogeneous_4_nodes_protocol}
\end{figure}
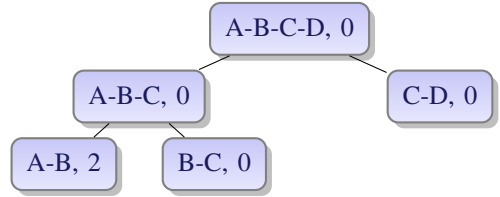

The protocol that maximizes the secret-key rate is presented in Fig.~\ref{fig:heterogeneous_4_nodes_protocol}.
Performing entanglement swapping first on the segment A-B-C, and subsequently with the elementary link C-D is advantageous with respect to the opposite tree shape.
This is because A-B will be (generally) generated before the other two links, as its value for $p_\text{gen}$ is higher. Thus, once generated, this link would decohere for long if swapped last.  

It is optimal to distill A-B link (twice), as it balances the waiting time needed for generating its adjacent B-C before the swapping, while also enhancing the quality of A-B.

\subsection{\textit{Faulty} Chains}\label{sec:results_faulty}

\begin{figure}[tb]
  \centering
  \includegraphics[width=\linewidth]{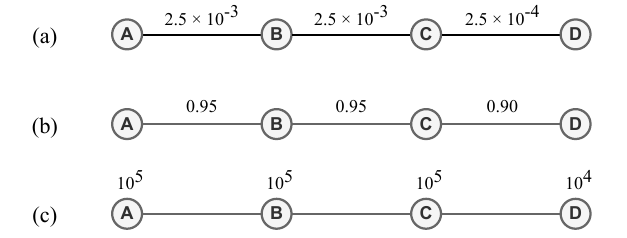}
  \caption{Visual representations of three faulty chains. In (a) and (b) respectively, values for $p_\text{gen}$ and $w_0$ are shown on top of the links. In (c), values for $t_\text{coh}$ are presented on top of the nodes. All the hardware parameters except for the ones presented are homogeneous.}
  \label{fig:faulty_chains}
\end{figure}

Finally, we consider three chains of $N=4$ nodes, homogeneous in their hardware, except for one \textit{faulty} node (or link) of lower quality.
A visual representation is shown in Fig.~\ref{fig:faulty_chains}.

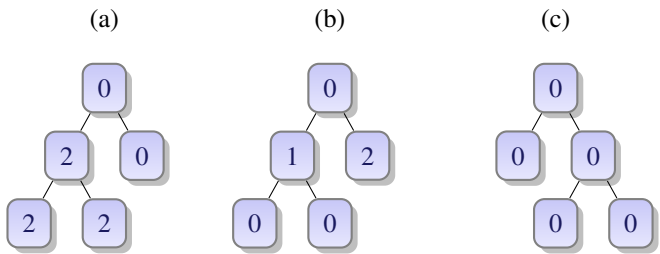
\begin{figure}[tb]
    \centering
    \begin{tikzpicture}[
        level distance=0.9cm,
        sibling distance=3cm,
        level 2/.style={sibling distance=1cm},
        every node/.style={inner sep=2mm},
        scale=1, transform shape
      ]

      \node at (-3, 0) {(a)}
      child {node[treenode] {0}
        child {node[treenode] {2}
          child {node[treenode] {2}}
          child {node[treenode] {2}}}
        child {node[treenode] {0}};
      };
      
      \node at (0, 0) {(b)}
      child {node[treenode] {0}
        child {node[treenode] {1}
          child {node[treenode] {0}}
          child {node[treenode] {0}}}
        child {node[treenode] {2}};
      };

      \node at (3, 0) {(c)}
      child {node[treenode] {0}
        child {node[treenode] {0}}
        child {node[treenode] {0}
          child {node[treenode] {0}}
          child {node[treenode] {0}}};
      };
    \end{tikzpicture}
    \caption{Results of the brute-force simulations on the chains shown in Fig.~\ref{fig:faulty_chains}.}
    \label{fig:faulty_chains_trees}
\end{figure}

For both the three chains the space of protocols is brute-forced, and the results are shown in Fig.~\ref{fig:faulty_chains_trees}.
In (a), where $p_\text{gen}$ is lower only for one out of the three links, it is optimal to first distill and swap the two links of higher quality, as it balances the waiting times while enhancing the quality of the entanglement, as observed in Section~\ref{sec:results_heterogeneous}.
In (b), the initial Werner parameter $w_0$ of the rightmost link (leaf) is lower than the other two. Thus, the optimal protocol distills the low-quality link while swapping the high-quality ones. 
In (c), the rightmost node has a lower quality for the time of coherence of its memory. Here, the link including this node is swapped first to decrease the impact of the decoherence.

\section{Conclusions}

In this work, we investigated asymmetric repeater chains, considering homogeneous and heterogeneous hardware settings. Our findings address the questions posed in Section~\ref{sec:introduction}.

For regimes in which elementary links have low quality, distillation significantly boosts the secret-key rate of the end-to-end link.
On the other hand, if the initial quality of the generated entangled pairs is already high, distillation becomes less beneficial, as the improvement does not justify the additional waiting time introduced by the protocol. 

For homogeneous chains, when distillation is not detrimental, performing it before the first entanglement swapping is generally optimal. 
The optimal number of rounds of distillation depends on the length of the repeater protocol, i.e. the distance at which we want to distribute entanglement. Generally, the larger the number of entanglement swapping instances, the larger the optimal number of distillation rounds.

In heterogeneous chains, it is advantageous to distill lower-quality links and swap earlier links with higher generation probabilities or connecting nodes with shorter coherence times. 
On the other hand, links that take longer to be generated should be swapped last, to balance the time at which the links are ready for later stages of the protocol.

Regarding the optimization approach, our results validate that Bayesian optimization effectively identifies optimal (or near-optimal) protocols in a fixed number of evaluations, offering a practical alternative to brute-force searches across all possible configurations.

\section*{Acknowledgment}

The authors thank support from JST Moonshot R\&D Grant No. JPMJMS226C. 

\end{document}